June 4, 2015

# X-ray Radiation Hardness of Fully-Depleted SOI MOSFETs and Its Improvement

Ikuo Kurachi[1], Kazuo Kobayashi[2], Hiroki Kasai[3], Marie Mochizuki[4], Masao Okihara[4],
Takaki Hatsui[2], Kazuhiko Hara[5], and Yasuo Arai[1]

[1]*High Energy Accelerator Research Organization (KEK), Ibaraki, Japan*
[2]*Riken SPring-8 Center, Hyogo, Japan*
[3]*Lapis Semiconductor Miyagi Co., Ltd., Miyagi, Japan*
[4]*Lapis Semiconductor Ltd., Kanagawa, Japan*
[5]*Faculty of Pure and Applied Science, University of Tsukuba, Ibaraki, Japan*

X-ray radiation hardness of FD-SOI n- and p-MOSFET has been investigated. After 1.4 kGy(Si) irradiation, 15% drain current increase for n-MOSFET and 20% drain current decrease for p-MOSFET are observed. From analysis of gmmax-Vsub, the major cause of n-MOSFET drain current change is the generated positive charge in BOX. On the other hand, the major cause of p-MOSFET drain current change is the radiation induced gate channel modulation by the generated positive charge in sidewall spacer. It is confirmed that the p-MOSFET drain current change is improved by higher PLDD dose. Thinner BOX is also proposed for further radiation hardness improvement.

PRESENTED AT

International Workshop on SOI Pixel Detector (SOIPIX2015)
Tohoku University, Sendai, Japan, 3-6, June, 2015

# 1  Introduction

Technical demand to utilize fully depleted Silicon-on Insulator (FD-SOI) devices in future large scale integrated circuit technology is increasing because of its scalability, ultra-low power consumption, and wide temperature range operation. In addition, the SOI structure is one of three dimensional devices if the active or passive elements are constructed in the handle wafer of SOI. Using this concepts, an X-ray image sensor as shown in Fig. 1 has been proposed and demonstrated [1]. Owing to monolithic structure of the SOI sensor, the fine pixel size can be realized which is indispensable for future sensors in science or medical application with high resolution. In this sensor, all of circuits are consisted in the FD-SOI layer. Therefore, radiation hardness of FD-SOI metal-oxide-semiconductor field-effect-transistors (MOSFET) has to be enough high to operate correctly in such circumstance. Actually, it is reported that the radiation hardness for single event upset (SEU) is improved by using FD-SOI technology, but total ionizing dose (TID) degradation is one of crucial issues [2]. In this presentation, the causes of X-ray irradiation damage of the FD-SOI n- and p-MOSFET are investigated in detail. Based on the investigation results, improvement methods for the radiation hardness are also proposed.

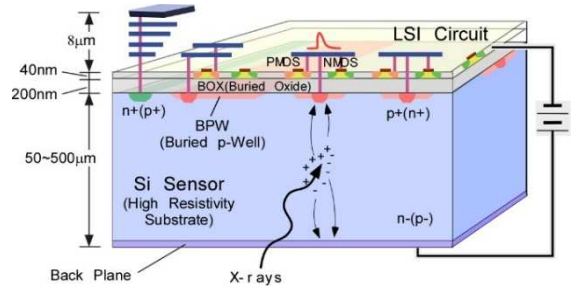

Figure 1. Schematic cross section of X-ray sensor

# 2  Experimental Procedure

A 0.2 μm FD-SOI CMOS process prepared by Lapis Semiconductor Co., Ltd. [3] is used to fabricate FD-SOI n- and p-MOSFETs. The thickness of buried oxide (BOX), SOI layer, and gate oxide are 200 nm, 40 nm, and 4.5 nm, respectively. A conventional lightly doped drain structure was employed to reduce electric field at the drain and short channel effect. Salicided diffusion and poly silicon was also adopted to make the parasitic resistance low and the reliable contact [4]. The X-ray irradiated and measured MOSFETs were designed 0.2 to 10 μm in gate length with constant gate width of 10 μm. $I_{ds}$-$V_{gs}$ curves with $|V_{ds}|$=0.1 V were measured before and after the X-ray irradiation. The linear region drain current, $I_{d\_lin}$, was measured at $|V_{ds}|$=0.1 V and $|V_{gs}|$=1.79 V and the threshold voltage, $V_{to}$, was extracted from extrapolation from $I_{ds}$-$V_{gs}$ around $g_{mmax}$. The X-ray irradiation was carried out by RIKEN using wafer-level X-ray irradiation system [5]. The dose rates were 0.018 Gy(Si)/s for analysis nd 3.0 Gy(Si) for confirmation experiments of the process improvement.

# 2    Results and Discussion

(1) TID radiation hardness of current FD-SOI MOSFETs

Figure 2 shows the linear region drain current change by the X-ray irradiation for n-MOSFET (Nch core) and p-MOSFET (Pch core). After 1.4 kGy(Si) irradiation, the change is around 15% increase for n-MOSFET and 20% decrease for p-MOSFET. Improvement of the radiation hardness has to be achieved at least more than 10-20 kGy(Si) within 10% drain current change. To improve the radiation hardness, the causes of drain current change by the X-ray irradiation are needed to find out. In general, MOSFET characteristic change by the X-ray irradiation is reported to be due to the positive charge generation in oxide and the interface state generation between oxide and silicon [6]. In FD-SOI case, there are two oxide films as BOX and gate oxide and two interfaces between gate oxide and SOI and between BOX and SOI. We should pay attention to these two oxide films and two interfaces. In the case of n-MOSFET, the drain current increases with the X-ray irradiation. Then, this may be caused by the positive change in gate oxide or BOX. To improve the radiation hardness of n-MOSFET, we have to distinguish which is the dominant cause of drain current change in n-MOSFET, charge in gate oxide or charge in BOX. In the case of p-MOSFET, the drain current decreases. Then, the suspected causes are (i)$|V_{to}|$ increase by the charge in gate oxide, (ii)positive back-bias by the charge in BOX, (iii)local $|V_{to}|$ increase by the charge in sidewall spacer, and (iv)mobility reduction by the generated interface states. Then, we need new analysis tool to identify which is the dominant cause in the p-MOSFET drain current change.

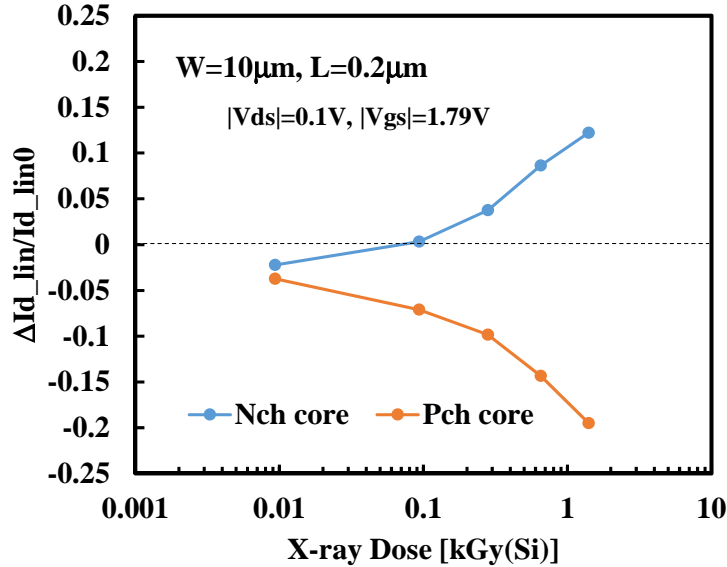

Figure 2. Drain current change for n- and p-MOSFET as a function of X-ray dose

(2) Cause of n-MOSFET drain current change

The possible causes to increase n-MOSFET drain current are the positive charge in gate oxide or BOX.

To distinguish the causes, we proposed analysis using maximum trans-conductance dependence on substrate bias ($g_{mmax}$-$V_{sub}$). The typical $g_{mmax}$-$V_{sub}$ characteristic is shown as blue line in Figure 3. After the X-ray irradiation, the $g_{mmax}$-$V_{sub}$ curve may shift as orange line in the figure.

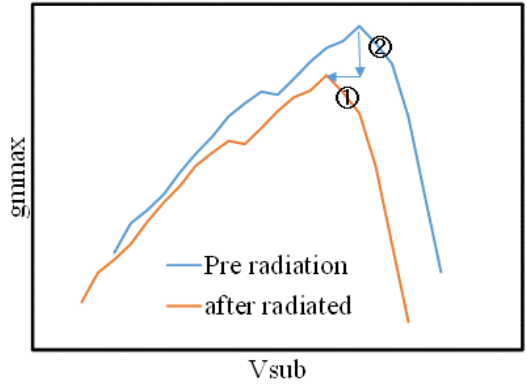

Figure 3. Schematic $g_{mmax}$-$V_{sub}$ characteristics before and after X-ray irradiation.

Then, the shift of ① is $\Delta V_{sub}$ and number of trapped holes in BOX, $N_{ot\_box}$, is given by

$$N_{ot\_box} = \frac{\epsilon}{qT_{ox\_box}} \Delta V_{sub} \quad (1)$$

where $T_{ox\_box}$ is the BOX oxide thickness, q is the elementary electron charge, and ε is the permittivity of oxide. The number of trapped holes in gate oxide $N_{ot\_gox}$ is also calculated from

$$N_{ot\_gox} = \frac{\epsilon}{qT_{ox\_gox}} \left[V_{tt\_deg}(0 + \Delta V_{sub}) - V_{tt\_int}(0)\right] \quad (2)$$

where $V_{tt\_int}(x)$ is the threshold voltage at $V_{sub}$=x for the fresh device and $V_{tt\_deg}(x)$ is that for the irradiated device. The $g_{mmax}$-$V_{sub}$ characteristics for each X-ray dose are shown in Figure 4. The calculated number of trapped holes in gate oxide and BOX is shown in Figure 5 (a) as a function of X-ray dose. Based on the results of Figure 5 (a), calculated $\Delta V_{to}$ is shown in Figure 5 (b). As shown in the figure, $V_{to}$ shift by charge in gate oxide is only 30 mV but that by charge in BOX is 130 mV. Therefore, the generated charge in BOX is the major cause of the drain current change in n-MOSFET. To improve the radiation hardness of n-MOSFET, the charge generation in BOX must be reduced. The proposal of improvement will be discussed later.

(3) Cause of p-MOSFET drain current change

As mentioned above, the possible causes of p-MOSFET drain current change are (i)|$V_{to}$| increase by the charge in gate oxide, (ii)positive back-bias by the charge in BOX, (iii)local |$V_{to}$| increase by the charge in sidewall spacer, and (iv)mobility reduction by the generated interface states. To distinguish which is the dominant cause, analysis by Terada's method [7] is used. In Terada's method, measured source to drain resistance $R_m$ including parasitic source drain resistance $R_{ext}$ is given by

$$R_m = \frac{\rho_{ch}}{W_{eff}}L + \left(\delta L \frac{\rho_{ch}}{W_{eff}} + R_{ext}\right) \quad (3)$$

Where $\rho_{ch}$ is the channel sheet resistance, $W_{eff}$ is the effective gate width, L is the gate length in design, $\delta L$ is the gate length bias from designed to effective gate length.

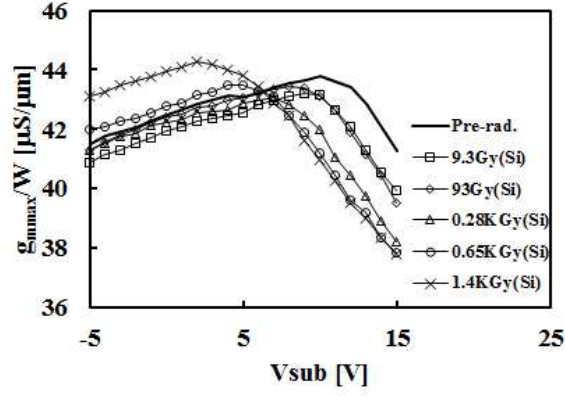

Figure 4. $g_{mmax}$-$V_{sub}$ characteristics for each X-ray dose.

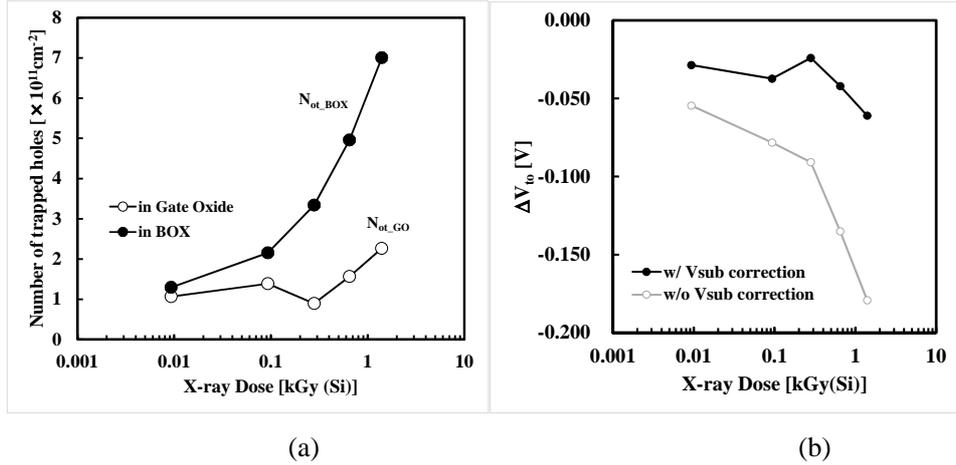

(a) (b)

Figure 5. (a) Number of trapped holes in gate oxide and BOX as a function of X-ray dose and (b) estimated $V_{to}$ shift from results of (a).

Using different gate length MOSFET set and varying $V_{gs}$-$V_{to}$, $\delta L$ and $R_{ext}$ can be extracted. In addition, $\rho_{ch}$ is given by

$$\rho_{ch} = \frac{1}{\mu C_{ox}(V_{gs} - V_{to})} \quad (4)$$

and the mobility μ can be calculated from the linear relationship between $1/\rho_{ch}$ and $V_{gs}$-$V_{to}$. The extracted change in μ, $\delta L$, and $R_{ext}$ is shown in Figure 6 as a function of X-ray dose both for n- and p-MOSFETs.

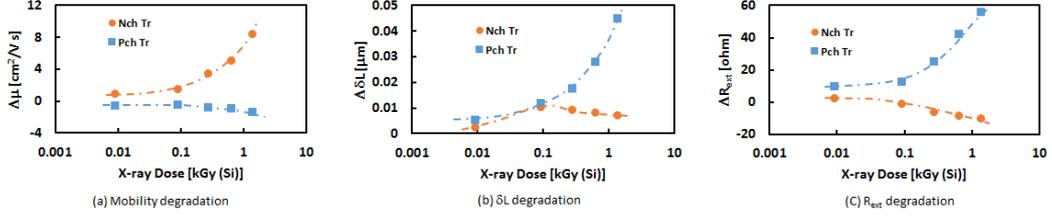

Figure. 6 Extracted μ, δL, and Rext by Terada's method as a function of X-ray dose.

In the case of n-MOSFET, only change of μ by the X-ray irradiation is observed whereas no change in δL and $R_{ext}$. This can be agreed with previous analysis of n-MOSFET radiation degradation. On the other hand, δL and $R_{ext}$ increase by the irradiation is observed in the case of p-MOSFET, whereas change of μ is small. Therefore, the dominant cause of p-MOSFET drain current change by X-ray irradiation should be related to the gate length modulation and the parasitic resistance increase. With consideration of these phenomena, the cause must be located at the edge of gate. We suspect the drain current change is due to the generated positive charge in sidewall spacer and the charge modulates the threshold voltage of the gate edge MOSFET and explained in [8]. In this explanation, the generated number of holes in the sidewall spacer $\Delta N_{ot\_sw}$ is a function of $\Delta\delta L$ as

$$\Delta N_{ot\_sw} = \frac{\epsilon}{2qL_{eff1}T_{ox\_sw}}\Delta\delta L \tag{5}$$

Where $L_{eff1}$ is the effective gate length of the gate edge MOSFET and $T_{ox\_sw}$ is the effective sidewall spacer thickness. From calculation of Eq. (5) using $\Delta\delta L$, $\Delta N_{ot\_sw}$ is estimated and compared to the generated number of holes in BOX as shown in Figure 7. We confirmed the same trend and the gate length modulation is due to the generated positive charge in sidewall spacer.

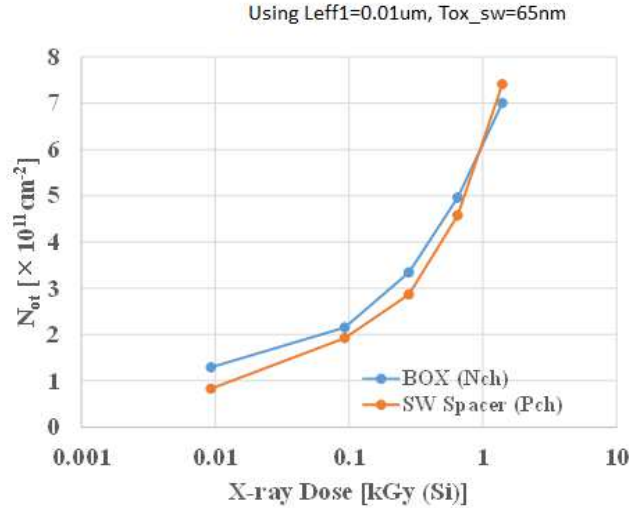

Figure 7. Generated number of holes in BOX and sidewall spacer.

(4) Improvement of p-MOSFET radiation hardness

The cause of the drain current change of p-MOSFET is the threshold voltage change at the edge of gate due to the generated positive charge in sidewall spacer. To suppress this effect, all channel of MOSFET must be controlled by gate potential not by the charge in sidewall spacer. Then, the LDD region must be totally overlapped by the gate and higher LDD dose is suspected to improve this effect. Figure 8 shows the drain current change ratios for different PLDD dose after 112 kGy(Si) X-ray irradiation. It is clear that higher PLDD dose such as 6 time or 10 time improves the ratios from 80% to 20%. This is really great improvement for the radiation hardness.

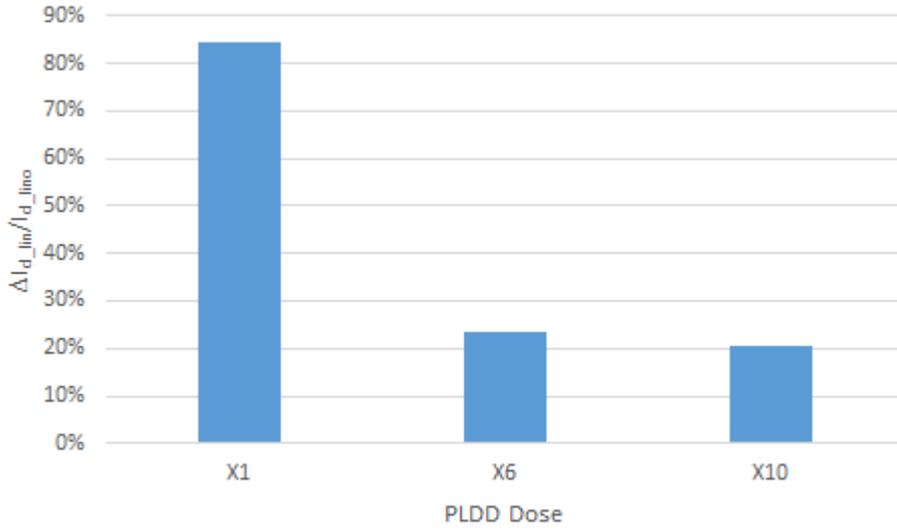

Figure 8 Drain current variation ratios for different PLDD dose.

(5) Proposal for further radiation hardness improvement

It is confirmed that higher PLDD dose greatly improves the radiation hardness but still there is current degradation after 112 kGy(Si) irradiation. In addition, we have to improve n-MOSFET radiation hardness. Based on measurement data, these degradations are caused by the generated positive charge in BOX. To reduce the effect by the charge in BOX, reducing BOX thickness is effective because the substrate bias shift by the generated positive charge is given by

$$\Delta V_{sub} = \frac{1}{\epsilon} \int_0^{t_{ox}} x\rho(x)\, dx \qquad (6)$$

where $\rho$ is the generated charge in BOX and $t_{ox}$ is the thickness of BOX. With assuming uniform charge distribution in BOX, $\Delta V_{sub}$ can be calculated for the different BOX thickness. The results are shown in Figure 9. When the BOX thickness is changed from 200 nm (current thickness) to 50 nm, the acceptable irradiation dose must be around three order different. If there is no high bias between handle silicon wafer and SOI, it is recommended to use the thinner BOX for the further improvement of radiation hardness.

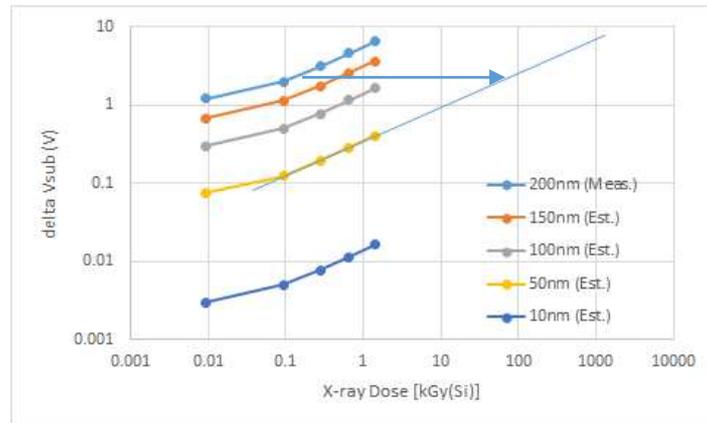

Figure 9. Estimated ΔVsub for different BOX thickness

## Summary

The FD-SOI n- and p-MOSFET degradation by X-ray irradiation has been investigated to find out the root cause of degradation. Based on investigated data, the suspected root cause is identified. In the case of n-MOSFET, the root cause is the positive charge generated by the X-ray irradiation. On the other hand, in the case of p-MOSFET, the root cause is the gate edge MOSFET threshold change due to the positive charge in sidewall spacer generated by the X-ray irradiation. The radiation hardness improvement has been done by higher PLDD dose based on the analysis results. Further improvement method, which is the thinner BOX, is also proposed.